\documentclass{aastex631}
\usepackage{hyperref}       % hyperlinks
\usepackage{url}            % simple URL typesetting
\usepackage{booktabs}       % professional-quality tables
\usepackage{amsfonts}       % blackboard math symbols
\usepackage{nicefrac}       % compact symbols for 1/2, etc.
\usepackage{microtype}      % microtypography
\usepackage{lipsum}		% Can be removed after putting your text content
\usepackage{graphicx}
\usepackage{natbib}
\usepackage{amsmath}
\usepackage{float}
\usepackage{tabularx}
\usepackage{xcolor}
\usepackage{graphicx}

\usepackage{hyperref}

\begin{document}

\title{An Interpretable AI Framework to Disentangle Self-Interacting and Cold Dark Matter in Galaxy Clusters: The CKAN Approach}

\correspondingauthor{Zhiyong Liu}
\email{liuzhy@xao.ac.cn}

\author[0009-0009-8188-5632]{Zhenyang Huang}
\affiliation{
Xinjiang Astronomical Observatory, Chinese Academy of Sciences, Urumqi 830011, China
}
\affiliation{
School of Astronomy and Space Science, University of Chinese Academy of Sciences, Beijing 101408, China}
% \email{huangzhenyang@xao.ac.cn}

\author[0009-0007-9418-2632]{Haihao Shi}
\affiliation{
Xinjiang Astronomical Observatory, Chinese Academy of Sciences, Urumqi 830011, China
}
\affiliation{
School of Astronomy and Space Science, University of Chinese Academy of Sciences, Beijing 101408, China}
% \email{shihaihao@xao.ac.cn}

\author[0000-0002-2136-1708]{Zhiyong Liu}
\affiliation{
Xinjiang Astronomical Observatory, Chinese Academy of Sciences, Urumqi 830011, China
}

\affiliation{
Key Laboratory of Radio Astronomy and Technology (Chinese Academy of Sciences), A20 Datun Road, Chaoyang District, Beijing, 100101, P. R. China}

\affiliation{
Key Laboratory of Xinjiang Radio Astrophysics, Urumqi 830011, China
}
% \email{liuzhy@xao.ac.cn}

\author[0000-0002-9786-8548]{Na Wang}
\affiliation{
Xinjiang Astronomical Observatory, Chinese Academy of Sciences, Urumqi 830011, China
}

\affiliation{
Key Laboratory of Radio Astronomy and Technology (Chinese Academy of Sciences), A20 Datun Road, Chaoyang District, Beijing, 100101, P. R. China}

\affiliation{
Key Laboratory of Xinjiang Radio Astrophysics, Urumqi 830011, China
}
% \email{na.wang@xao.ac.cn}

% \collaboration{20}{(AAS Journals Data Editors)}

% \author{F.X Timmes}
% \affiliation{Arizona State University}
% \affiliation{AAS Journals Associate Editor-in-Chief}

% \author{Amy Hendrickson}
% \altaffiliation{AASTeX v6+ programmer}
% \affiliation{TeXnology Inc.}

% \author{Julie Steffen}
% \affiliation{AAS Director of Publishing}
% \affiliation{American Astronomical Society \\
% 1667 K Street NW, Suite 800 \\
% Washington, DC 20006, USA}




\begin{abstract}

Convolutional neural networks have shown their ability to differentiate between self-interacting dark matter (SIDM) and cold dark matter (CDM) on galaxy cluster scales. 
However, their large parameter counts and “black-box” nature make it difficult to assess whether their decisions adhere to physical principles.
% However, they are limited by their large number of parameters and "black-box" nature, making it difficult to understand the basis of their outputs and whether they are consistent with physical principles. 
% To address this issue, we build a Convolutional Kolmogorov-Arnold Network (CKAN), which uses fewer parameters and offers greater interpretability to efficiently disentangle these two dark matter models, and propose a novel analytical approach to better understand how the network making decisions. 
To address this issue, we have built a Convolutional Kolmogorov–Arnold Network (CKAN) that reduces parameter count and enhances interpretability, and propose a novel analytical framework to understand the network’s decision‐making process.
With this framework, we leverage our network to qualitatively assess the offset between the dark matter distribution center and the galaxy cluster center, as well as the size of heating regions in different models. 
These findings are consistent with current theoretical predictions and show the reliability and interpretability of our network. 
By combining network interpretability with unseen test results, we also estimate that for SIDM in galaxy clusters, the minimum cross-section $(\sigma/m)_{\mathrm{th}}$ required to reliably identify its collisional nature falls between $0.1\,\mathrm{cm}^2/\mathrm{g}$ and $0.3\,\mathrm{cm}^2/\mathrm{g}$.
% We also estimate that the minimum SIDM cross-section $(\sigma/m)_{th}$ required to disentangle SIDM from CDM in galaxy clusters is between $0.1\,\mathrm{cm}^2/\mathrm{g}$ and $0.3\,\mathrm{cm}^2/\mathrm{g}$ by combining network interpretability with unseen test results.
Moreover, CKAN maintains robust performance under simulated JWST and Euclid noise, highlighting its promise for application to forthcoming observational surveys.

\end{abstract}



\section{Introduction} \label{sec:intro}

Although dark matter accounts for a significant 27\% of the universe’s total energy density \citep{2020A&A...641A...6P}, its specific nature remains one of the most elusive questions in astrophysics, with scientists continuously striving to uncover its properties since it was first introduced. 
The nature of dark matter remains uncertain, prompting physicists to propose numerous candidates to explain its properties, such as axions \citep{Weinberg,Caputo}, fuzzy dark matter \citep{Hu:2000ke,Sipple:2024svt}, primordial black holes \citep{Carr:1974nx,Green:2024bam} and Weakly Interacting Massive Particles (WIMPs) \citep{Spergel1985,XENON:2020kmp}.
Among them, Cold Dark Matter (CDM) \citep{Blumenthal:1984bp,Spergel_2003,Drukier:1986tm} and Self-Interacting Dark Matter (SIDM) \citep{Spergel:1999mh,Zavala:2012us,Tulin_YuHai-Bo} are two of the widely studied candidates. 
While CDM has been effective in explaining large-scale phenomena like quasar evolution and the cosmic microwave background \citep{2016A&A...594A..13P, 2005Natur.435..629S, 2011ApJS..192...18K}, it has faced difficulties in addressing recent small-scale observations, which reveal differences that CDM struggles to explain \textcolor{black}{compared to SIDM}, \textcolor{black}{such as the core-cusp problem} \citep{1994Natur.370..629M,2011ApJ...742...20W,Weinberg_2015}, \textcolor{black}{and the excess of small-scale gravitational lenses in galaxy clusters} \citep{2020Sci...369.1347M}. 
Alternatively, SIDM not only offers a more satisfactory explanation for strong lensing effects in galaxy clusters \citep{2020Sci...369.1347M, 2024ApJ...970..143T, 2023A&A...678L...2M}, but also addresses key issues such as the diverse rotation curves of dwarf galaxies and the soft core problem in galaxy centers \citep{Marchesini_2002, 10.1046/j.1365-8711.2000.03555.x, 1994ApJ...427L...1F}.

Galaxy clusters have long been among the most ideal laboratories for scrutinizing and refining theories of dark matter.
Their massive concentrations can heighten many subtle changes in the particle physics model, altering formation and structure at an observable scale \citep{Kalhoefer2014, 2019MNRAS.488.3646R}. 
Moreover, their abundance means that dark matter self-interaction can be studied statistically over many objects using gravitational lensing, which does not assume any dynamical state \citep{2015Sci...347.1462H}.
In galaxy clusters, the subtle differences between SIDM and CDM manifest most prominently on small scales.
In SIDM halos, repeated dark matter particle collisions lead to heat transfer from the hot outer regions into the cooler core, producing a characteristic central density “core” and rounder isodensity contours \citep{10.1093/mnras/sts535} compared to the steep, triaxial cusps of a Navarro–Frenk–White profile expected in CDM-only scenarios \citep{1996ApJ...462..563N}. 
In merging galaxy clusters, the finite self-interaction cross-section of SIDM can induce measurable offsets between the total mass peaks and the centroids of collisionless galaxies offsets \citep{2008ApJ...679.1173R}. 
Notably, such peak offsets depend on various parameters of both dark matter and the host clusters, such as the strength of self-interactions between dark matter particles \citep{Kalhoefer2014}, the cross-section, cluster mass, and the amount of stellar mass \citep{2019MNRAS.488.1572H}.
These peak offsets thus provide one of the most direct probes of the SIDM cross-section \citep{2015Sci...347.1462H}. 
However, cosmological hydrodynamical simulations (the BAHAMAS–SIDM suite) have demonstrated that baryonic processes, including asymmetric gas distributions and stellar mass centroids, can induce offsets of a few kiloparsecs in pure CDM halos, thereby closely mimicking the SIDM signal at low cross-sections and complicating the interpretation of observed separations \citep{2019MNRAS.488.3646R}.

Over the past decade, numerous observables have been proposed to probe dark matter, including measurements of halo shapes \citep{10.1086/324138,2021MNRAS.500.2627H}, the abundance of strong lensing arcs \citep{Meneghetti2000GiantCA,2019MNRAS.488.3646R}, wobbles of the brightest cluster galaxy \citep{2017MNRAS.469.1414K,10.1093/mnras/stx2084}, trails within dark matter halos \citep{10.1093/mnras/stx855,2017MNRAS.464.3991H}, peak offsets in merging clusters \citep{2004ApJ...606..819M,2013MNRAS.433.1517H}, and mass loss during cluster mergers \citep{2008ApJ...679.1173R}.
However, they are often equally sensitive to baryonic feedback, making them degenerate with collisionless dark matter \citep{2019MNRAS.488.3646R,2023MNRAS.521.3172R}.
In a bid to find a generalized method that encapsulates all these different observables, convolutional neural networks (CNNs), enhanced by classical image-processing techniques from machine learning, have been introduced to disentangle self-interacting from collisionless dark matter and have proven to be an efficient method for isolating the subtle signatures of SIDM within complex baryonic phenomena \citep{harvey2024}.

While many deep learning networks have been highly successful, they are often parameter-intensive and lack transparency, making their internal mechanisms difficult to interpret.
This challenge has remained a persistent issue in the AI-for-science domain \citep{nature}. 
Effectively addressing this challenge requires moving beyond methods typically deemed “black-box” predictors, because a comprehensive understanding of the phenomena demands interpretable models \citep{whitepaper}. 
Developing interpretable networks for data harboring different dark matter model information not only enables efficient data processing and analysis, but also holds the potential to uncover the subtle disentangling features among different dark matter models through network interpretability and generate novel hypotheses or insights.

Recently, Kolmogorov-Arnold Networks (KANs), inspired by the Kolmogorov-Arnold representation theorem and introducing learnable activation functions that significantly enhance model expressiveness while preserving interpretability, have emerged as a promising solution to this challenge \citep{liu2024kankolmogorovarnoldnetworks}. 
The Convolutional Kolmogorov-Arnold Network (CKAN) extends this innovation to convolutional layers \citep{bodner2024convolutionalkolmogorovarnoldnetworks}. 
The emergence of this new network allows us to remodel some astronomical data and further explore the physical features hidden in the data through the interpretability of the network.

This study aims to develop a neural network, CKAN, with fewer parameters and enhanced interpretability, capable of effectively disentangling between SIDM and CDM in galaxy clusters while extracting relevant physical features from the input data. 
We present an interpretable neural-network–driven framework to infer the features of dark matter interactions and demonstrate its potential for application in forthcoming observational surveys.
In this paper, we introduce the dataset and explain CKAN network architecture, along with its results and generalization ability in Section \ref{sec:Methods}.  
Section \ref{sec:Physical Significance of the CKAN} focuses on analyzing the interpretability of the CKAN, and Section \ref{sec:4} explores its robustness. Lastly, we make a summary in Section \ref{sec:conclusion}.

\section{Methods}\label{sec:Methods}

\subsection{Datasets}

The performance of deep learning models is highly dependent on the quality of the input data. For this reason, we utilize input data from the BAHAMAS-SIDM simulations, based on the Gadget-3 hydrodynamical algorithm \citep{10.1111/j.1365-2966.2005.09655.x}, which incorporate a wide range of complex mechanisms, making them ideal for deep learning applications. 
These simulations account for dark matter self-interactions and a comprehensive baryonic feedback mechanism, which includes radiative cooling \citep{10.1111/j.1365-2966.2008.14191.x}, AGN feedback \citep{DallaVecchia:2008ouv}, star formation \citep{Schaye:2007ss}, and stellar and chemical dynamics \citep{Wiersma:2009wf}. 
The stellar and AGN feedback parameters have been fine-tuned to match the known fraction of gas in clusters \citep{10.1093/mnras/stw2792, Mccarthy:2017yqf}. 
Each model is simulated within a 400 Mpc/h cosmological box containing $2 \times 1024^{3}$ particles, using WMAP9 cosmological parameters. 
The dark matter particle mass is $m_{\mathrm{DM}} = 5.5 \times 10^{9} \,\text{M}_{\odot}$, and the initial gas particle mass is $m_{\mathrm{gas}} = 1.1 \times 10^{9} \,\text{M}_{\odot}$. 
For redshifts below $z=3$, the Plummer-equivalent gravitational softening length is 5.7 kpc \citep{Bennett_2013}.

The dark matter self-interactions are elastic, velocity-independent and isotropic. At each time step, the nearest particles (using a fixed radius) are found, and using a probabilistic recipe, each particle is scattered, conserving momentum and energy.
For a comprehensive description of the SIDM model used and its implementation, see \citep{2019MNRAS.488.3646R}.
The simulation data includes three collisionless dark matter models with varying levels of AGN feedback (CDM-low AGN, CDM fiducial AGN, and CDM-hi AGN), as well as three SIDM models with cross-sections of $\sigma_{\mathrm{DM}}/m = 0.1, 0.3, 1.0\,\mathrm{cm}^2/\mathrm{g}$, and consists of three channels: 2D distribution maps of total mass, stellar mass, and X-ray emissions.
More specific details and sources about the dataset can be found in \citep{harvey2024}.

\subsection{Kolmogorov-Arnold Networks}

The Kolmogorov-Arnold representation theorem states that if $f$ is a multivariate continuous function on a bounded domain, then $f$ can be written as a finite composition of continuous functions of a single variable and the binary operation of addition. More specifically, for a smooth $f$ : $[0, 1]^n \rightarrow \mathbb{R}$:

\begin{equation}\color{black}
f(x_1, \ldots, x_n) = \sum_{q=1}^{2n+1} \Phi_q \left( \sum_{p=1}^n \phi_{q,p}(x_p) \right).
\label{kanRT}
\end{equation}

Similar to Multilayer Perceptrons (MLPs), which are based on the Universal Approximation Theorem, KANs integrate the Kolmogorov-Arnold representation theorem into neural networks. 
Instead of relying on fixed activation functions between nodes, KANs employ activation functions parameterized by trainable B-splines, making them learnable during training. 
As a result, both \( \Phi_q \) and \( \phi_{q,p} \) in \autoref{kanRT} become functions that the network can adaptively learn.
Specifically, at each layer, the transformation \( \Phi_l \) operates on the input \( x_l \) to produce the input for the next layer, \( x_{l+1} \), as represented in the following matrix:

\begin{equation}\color{black}
    {x}_{l+1} = \Phi_l(x_l) =
    \begin{pmatrix}
        \phi_{l,1,1}(\cdot) & \phi_{l,1,2}(\cdot) & \cdots & \phi_{l,1,n_{l}}(\cdot) \\
        \phi_{l,2,1}(\cdot) & \phi_{l,2,2}(\cdot) & \cdots & \phi_{l,2,n_{l}}(\cdot) \\
        \vdots & \vdots & & \vdots \\
        \phi_{l,n_{l+1},1}(\cdot) & \phi_{l,n_{l+1},2}(\cdot) & \cdots & \phi_{l,n_{l+1},n_{l}}(\cdot) \\
    \end{pmatrix}\label{eq:kanforwardmatrix}
    {x}_{l}.
\end{equation}

The final output of the network is given by the composition of all layer transformations, as shown in the following equation:
\begin{equation}\color{black}
    \text{KANs}(x) = (\Phi_{L-1} \circ \Phi_{L-2} \circ \cdots \circ \Phi_0)(x).
\end{equation}

We can also rewrite the above equation to make it more analogous to \autoref{kanRT}, assuming output dimension $n_{L}=1$, and define $f({x})\equiv {\rm KAN}({x})$:

\begin{equation}\color{black}
    f({x})=\sum_{i_{L-1}=1}^{n_{L-1}}\phi_{L-1,i_{L},i_{L-1}}\left(\sum_{i_{L-2}=1}^{n_{L-2}}\cdots\left(\sum_{i_2=1}^{n_2}\phi_{2,i_3,i_2}\left(\sum_{i_1=1}^{n_1}\phi_{1,i_2,i_1}\left(\sum_{i_0=1}^{n_0}\phi_{0,i_1,i_0}(x_{i_0})\right)\right)\right)\cdots\right).
\end{equation}

Compared to MLPs with fixed activation functions, these diverse set of \( \{\phi_{l,j,k}\} \) not only enables the extraction of richer nonlinear features from the data through training but can also be further fitted to elementary functions, which enables KAN-based networks to achieve the same level of accuracy with fewer nodes and parameters, thereby enhancing interpretability \citep{liu2024kankolmogorovarnoldnetworks}.

Specifically, in all experiments reported in this paper, the final network is selected by retaining the checkpoint exhibiting the lowest validation loss throughout training.
Subsequently, node-level functional approximations are carried out using KAN’s built-in symbolic regression library, which demonstrates exceptional fidelity for all activated nodes (coefficient of determination \( R^2 > 0.99 \)). In this context, \( R^2 \) is defined as follows:

\textcolor{black}{\begin{equation}
      R^2 = 1 - \frac{SS_{\text{res}}}{SS_{\text{tot}}} = 1 - \frac{\sum_{i=1}^n (y_i - \hat{y}_i)^2}{\sum_{i=1}^n (y_i - \bar{y})^2}.
\end{equation}}

\subsection{CKAN}

% In this section, we introduce the CKAN architecture for this study \citep{huangzhenyang_2024_14174520} and present its training performance.
In this section, we introduce the CKAN architecture for this study \citep{huangzhenyang_2024_14174520} and present its training performance.
CKAN combines the strengths of KAN and CNN, consisting of convolutional kernels based on the Kolmogorov-Arnold representation theorem and fully connected KAN-layers. 
We test a series of CKAN variants spanning a range of parameter counts and layer depths, and identify the architecture with the fewest parameters and layers without sacrificing classification accuracy.

\begin{figure}[H]
    \centering
    \epsscale{0.80}
    \plotone{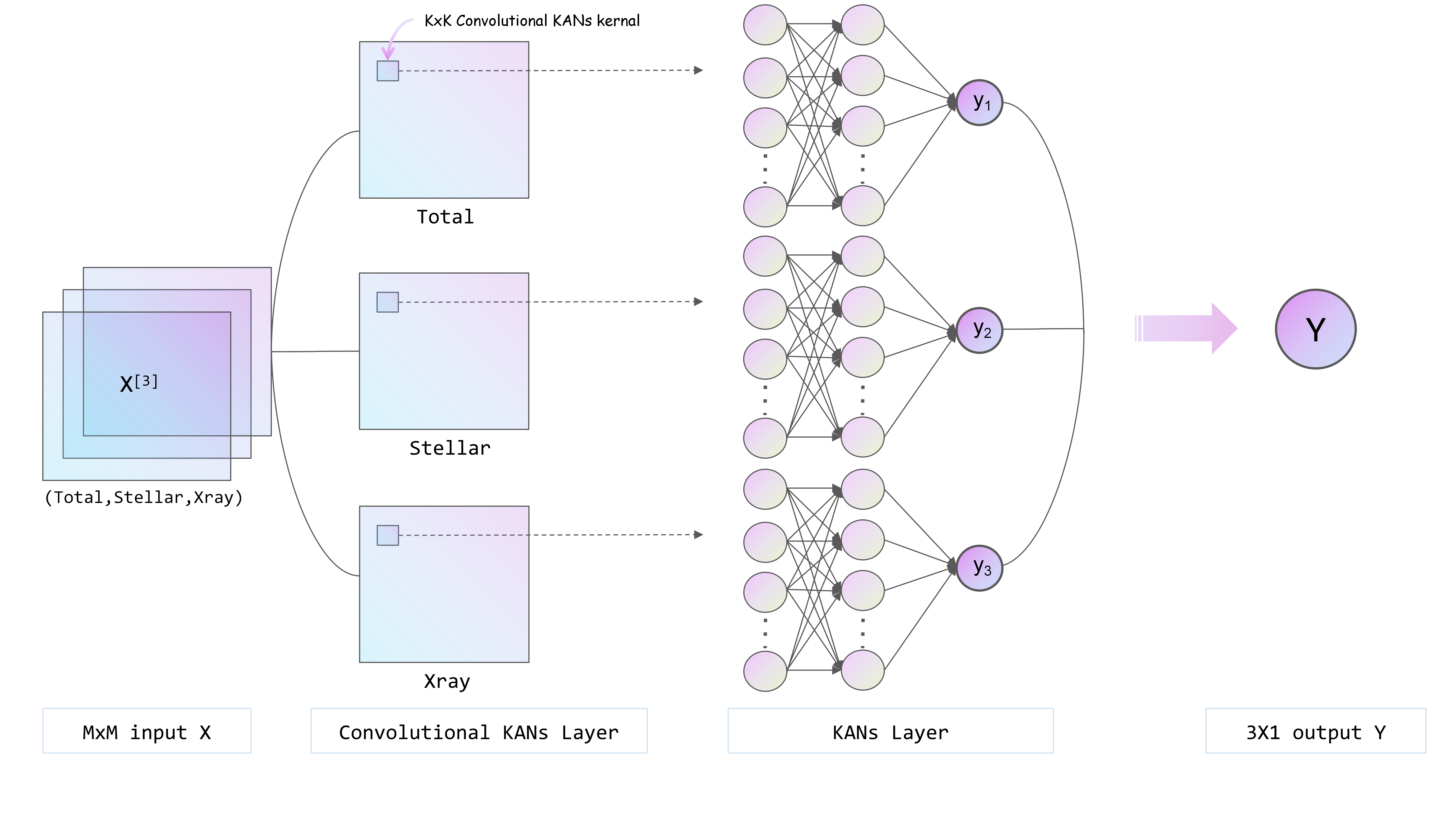}
    \caption{Schematic illustration of our network. To better study the features extracted by each channel, we feed the distribution maps from the three channels (total, stellar, and X-ray) into the convolutional KAN  kernel and fully connected layers, which share the same configuration, and then sum the results $y_{i}$ to produce the network's final output $Y$. The network's loss function is set to cross-entropy, with its output being a three-class vector corresponding to the probabilities of each class. } 
    \label{network}
\end{figure}

Figure \ref{network} shows the final CKAN architecture. 
Training CKAN produces three parallel subnetworks, one per input channel, that extract distinct features from each.
This allows us to gain clear insight into each channel’s decision-making process in identifying different dark matter models.

\begin{figure}[H]
    \centering
    \begin{minipage}[b]{0.48\textwidth}
        \centering
        \includegraphics[width=\textwidth]{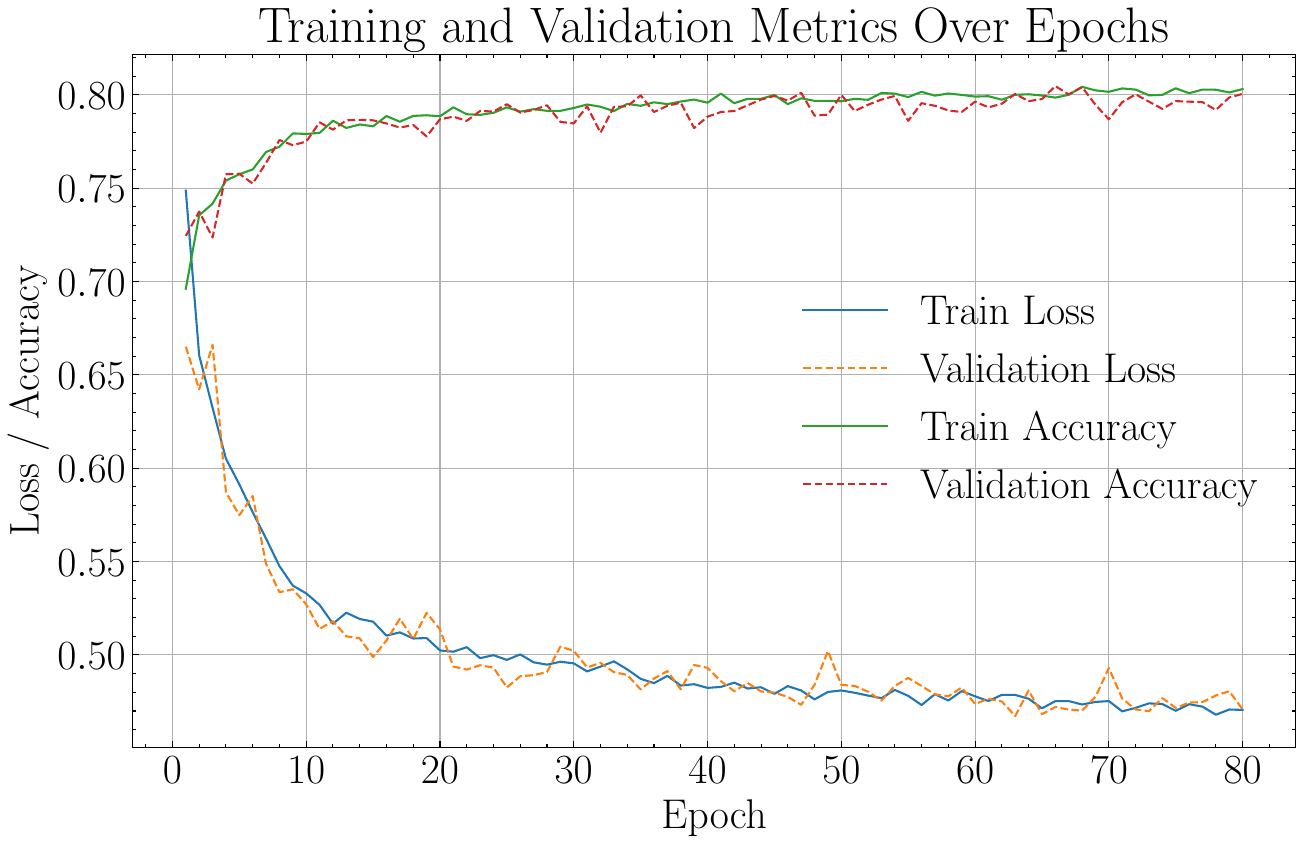}
        \caption{
        Training metrics of CKAN over 80 epochs. 
        The network shows no evidence of overfitting, either in terms of loss or accuracy.}
        \label{fig:result1}
    \end{minipage}
    \hspace{0.01\textwidth}
    \begin{minipage}[b]{0.48\textwidth}
        \centering
        \includegraphics[width=0.85\textwidth]{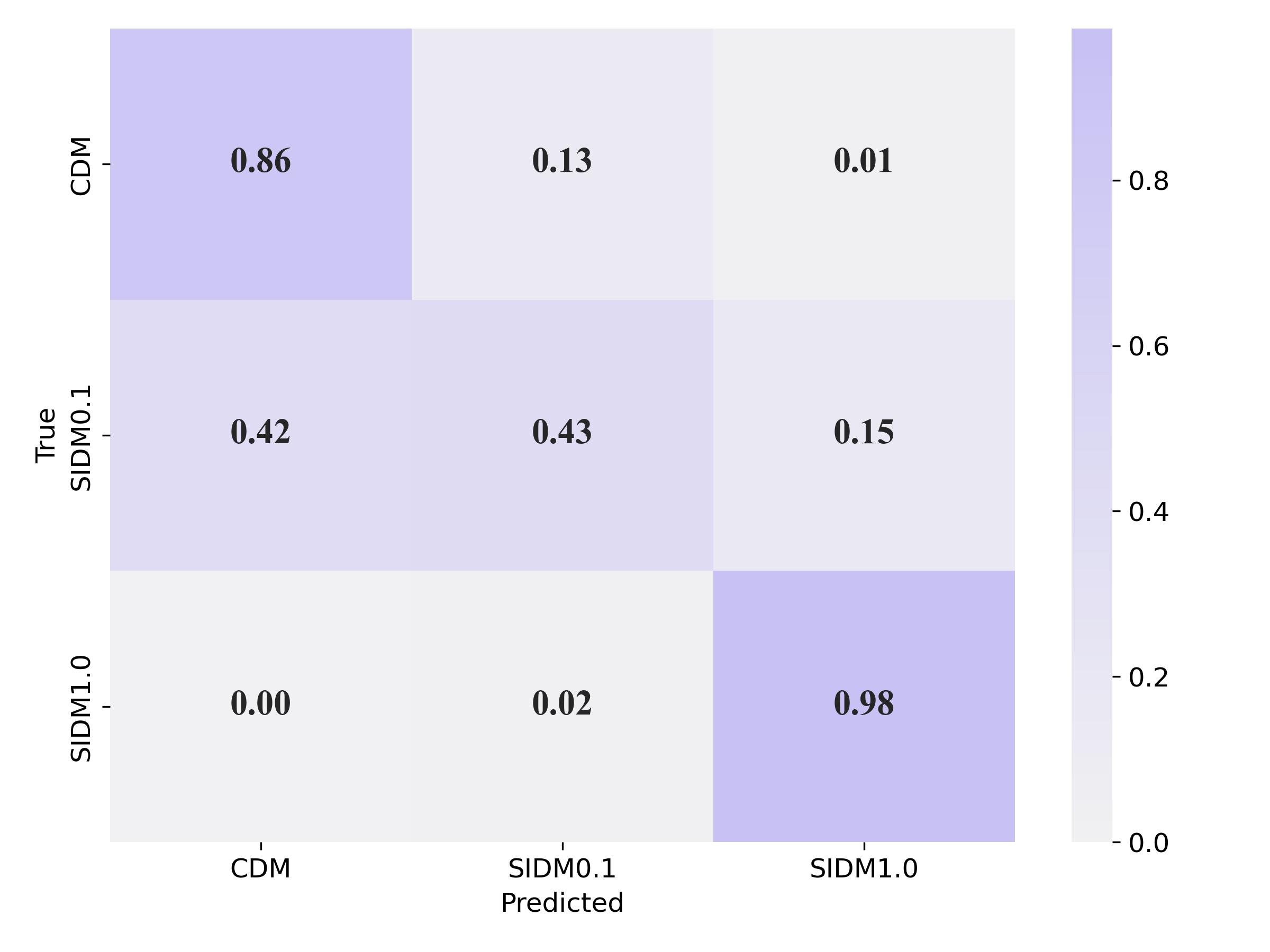}
        \caption{The confusion matrix of the network on the validation set; the CDM includes CDM-low AGN, CDM fiducial AGN, and CDM-hi AGN.}
        \label{fig:Confusion_Matrix}
    \end{minipage}
    \label{fig:combined_result}
\end{figure}

The training process is carried out using supervised learning on a dataset consisting of the CDM (including CDM-low AGN, CDM fiducial AGN, and CDM-hi AGN), SIDM0.1, and SIDM1.0 models, which were split into training and validation sets in an 80:20 ratio. 
To enhance the dataset, we applied data augmentation techniques, such as rotation, scaling, and contrast adjustments, following successful methods used in galaxy classification tasks \citep{10.1093/mnras/stv632}. 
We use Adam as the optimizer, with a learning rate of $10^{-3}$ and a batch size of 32.

After training CKAN with the aforementioned settings for 80 epochs, it achieves an accuracy of approximately 80\%, with the detailed training metrics shown in \autoref{fig:result1}. 
\autoref{fig:Confusion_Matrix} highlights CKAN's performance on the validation set, showing results similar to CNN \citep{harvey2024}, where the network performs well on CDM and SIDM1.0 data but struggles to clearly classify between CDM and SIDM0.1. 
However, CKAN’s interpretability allows us to further explore and explain this challenge, as discussed in Section \ref{sec:Physical Significance of the CKAN}. \autoref{table:params} shows a comparison of the parameters for different CKANs and CNNs, along with the results from training with both single-channel and dual-channel data. 
The best-performing single channel is the total, while the best dual-channel combination is total and X-ray.

\begin{deluxetable}{ccc}
\tablecaption{Parameter comparison of various CKANs and CNNs \citep{harvey2024} \label{table:params}}
\tablewidth{0pt}
\tablehead{
\colhead{Model} & \colhead{Parameters} & \colhead{Accuracy}
}
\startdata
InceptionCNN & 13,480,419 & $>80\%$ \\
CKAN & 14,208 & $80\%$ \\
CKAN-BigConvs & 23,250 & $80\%$ \\
SimpleCNN & 1,204,005 & $<80\%$ \\
CKAN-total+X-ray & 9,742 & $78\%$ \\
CKAN-onlytotal & 4,736 & $77\%$ \\
\enddata
% \tablecomments{An example of comparing model parameters and accuracies.}
\end{deluxetable}

\subsection{Generalization Ability of the CKAN}\label{Generalization}

Testing the neural network's ability to generalize is crucial for assessing its robustness and applicability beyond the training data. 
In this section, we evaluate the generalization performance of the trained CKAN on galaxy cluster data not included in the training set.

\begin{figure}[H]
    \centering
    \begin{minipage}[b]{0.48\textwidth}
        \centering
        \includegraphics[width=\textwidth]{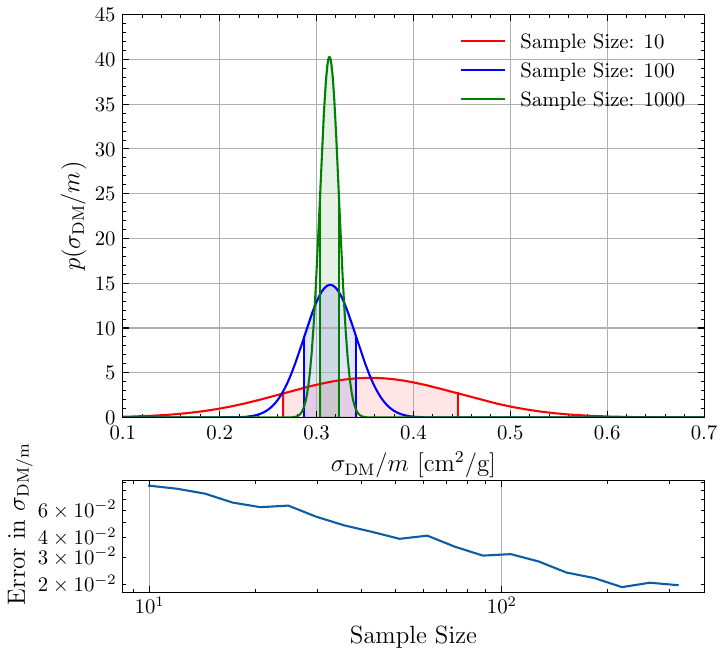}
    \end{minipage}
    \begin{minipage}[b]{0.48\textwidth}
        \centering
        \includegraphics[width=\textwidth]{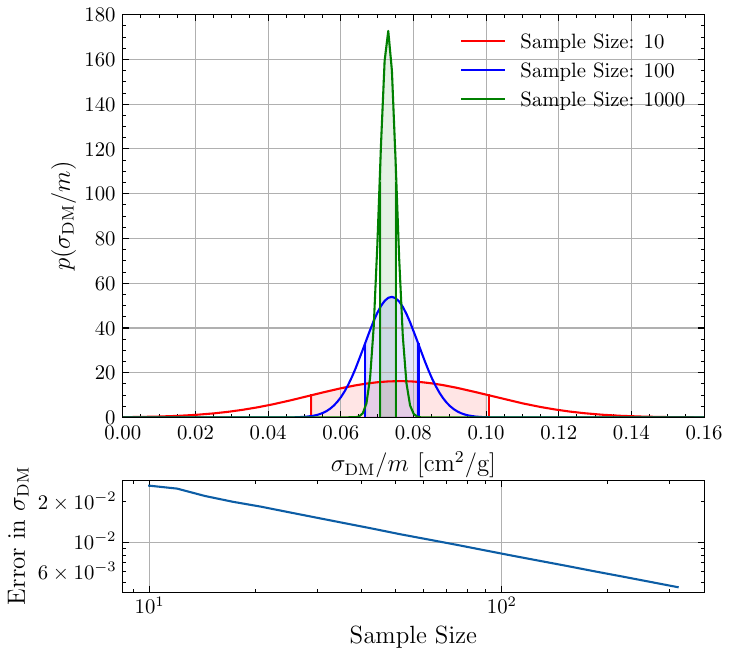}
    \end{minipage}
    \caption{
    Generalization performance of our CKAN on unseen clusters. 
    Left: training on CDM-low AGN, CDM fiducial AGN, CDM-hi AGN and two SIDM models ($\sigma_{\mathrm{DM}}/m = 0.1, 1\,\mathrm{cm}^2/\mathrm{g}$), and then testing on the unseen SIDM models ($\sigma_{\mathrm{DM}}/m = 0.3\, \mathrm{cm}^2/\mathrm{g}$).
    Right: training on CDM-low AGN, CDM fiducial AGN, CDM-hi AGN and two SIDM models ($\sigma_{\mathrm{DM}}/m = 0.3,1\,\mathrm{cm}^2/\mathrm{g}$), and then testing on the unseen SIDM models ($\sigma_{\mathrm{DM}}/m = 0.1\,\mathrm{cm}^2/\mathrm{g}$).}
    \label{fig:CKAN_blindtest_result}
\end{figure}

Firstly, the output of the CKAN can be understood as probabilities for estimating the scattering cross-sections of galaxy clusters that were not included in the training process. 
The detailed calculation process can be found in \autoref{calculate:pred}. 
Next, we evaluate CKAN on galaxy cluster data excluded from the training set, assessing its performance on sample sizes of 10, 100, and 1000 clusters.
\autoref{fig:CKAN_blindtest_result} demonstrates that CKAN provides accurate estimates for the SIDM0.3 test data; for ten galaxy cluster samples, the error in $\sigma_{\text{DM}}/m$ is on the order of $10^{-2}\,\mathrm{cm}^2/\mathrm{g}$, with the network outputting a prediction that slightly exceeds the true value. 
In contrast, for the SIDM0.1 test data, the error is even smaller, and the network produces a prediction that is marginally lower than the true value.

\section{Physical Significance of the CKAN}\label{sec:Physical Significance of the CKAN}

\begin{figure}[h]
    \centering
    \plotone{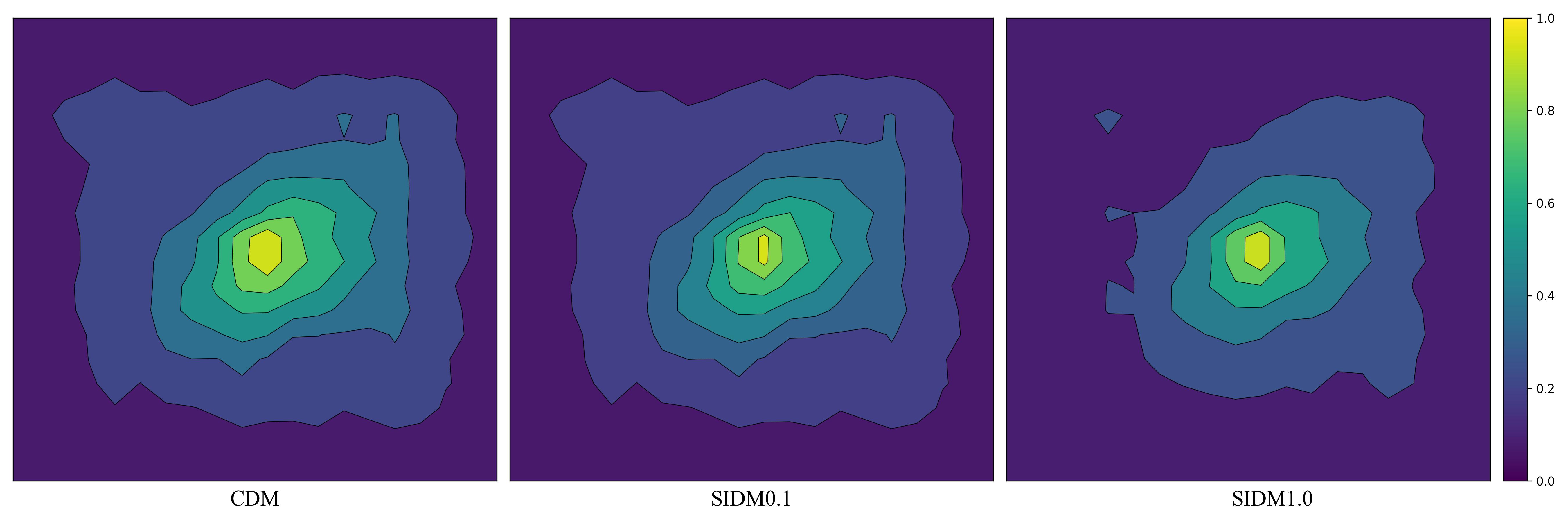}
    \caption{The feature contribution map for the classification of different dark matter models in the total channel approach. From left to right, the feature contribution maps obtained for the CDM, SIDM0.1, and SIDM1.0 models are shown. The color scale bar on the right indicates the normalized $h_i$ values corresponding to each point in the original distribution. Overlaid contours on each panel highlight the distribution pattern of $h_i$. Notably, both the intensity regions and overall morphology exhibit significant variations with changes in the self-interaction strength. }
    \label{figurescombined_total}
\end{figure}

\begin{figure}[h]
    \centering
    \plotone{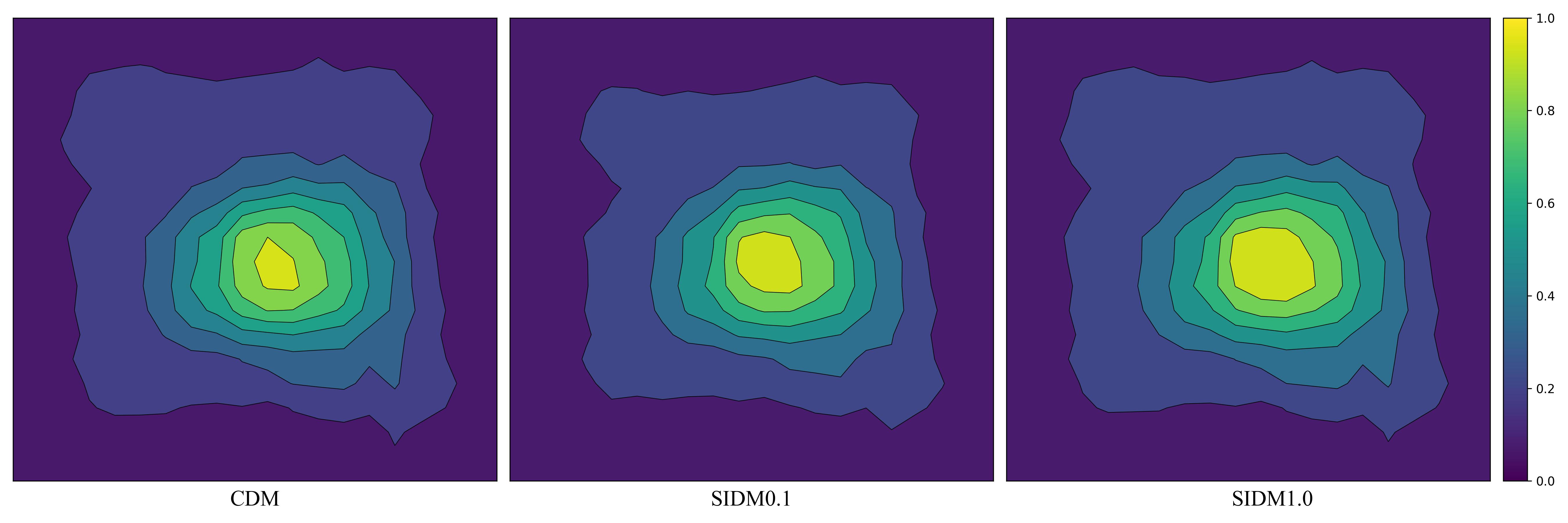}
    \caption{The feature contribution map for the classification of different dark matter models in the X-ray channel approach. From left to right, the feature contribution maps obtained for the CDM, SIDM0.1, and SIDM1.0 models are shown. The color scale bar on the right indicates the normalized $h_i$ values for each point in the original distribution. Overlaid contours in each panel emphasize the spatial pattern of $h_i$. Notably, both the intensity regions and overall morphology exhibit significant variations as the self-interaction strength changes.    
   }
    \label{figures:combined_xray}
\end{figure}

To probe CKAN’s interpretability and the principles behind its classifications, we apply the KAN function library to symbolically decompose each network component. 
This produces a contribution score \(h_i\) for every pixel in the input maps, indicating its influence on the output probability amplitudes (see \autoref{calculate:feature:map} for details). 
To reduce computational cost, we compute these scores via numerical differentiation and then average the results over multiple runs with different random seeds to mitigate stochasticity in the training process.
Finally, we compute feature scores for every pixel in the original distribution maps and average these scores for each model group, including CDM (with three AGN feedback variants), SIDM0.1, and SIDM1.0, which are shown in \autoref{figurescombined_total} and \autoref{figures:combined_xray}.  
These figures present heatmaps illustrating the contributions of the original distributions from the two most influential channels in the classification results (total and X-ray).

As shown in \autoref{figurescombined_total}, the total contribution map reveals an intriguing asymmetry, where the density on the upper-left side of the image center is higher than on the opposite side, with the gradient on this side following an exponential or inverse-square law, while the other side exhibits a more gradual, linear decline, a phenomenon that may be attributed to the misalignment between the dark matter core and the galaxy cluster center. 
Increasing evidence supports the existence of halo miscentering and mass bias in galaxy clusters \citep{2012ApJ...757....2G}. 
\cite{10.1093/mnras/stab1198} had noted that frequent interactions in merging clusters of similar mass can lead to significant offsets between the distributions of galaxies and the dark matter core.

As the collisional scale for the mean free path is inversely proportional to \(\sigma_{DM}\) \citep{Balberg_2002}, and self-interactions reduce the central density of dark matter \citep{2019MNRAS.488.3646R}, the frequency of dark matter collisions is directly proportional to \(\sigma_{DM}\). 
From \autoref{figurescombined_total}, we can conclude that a larger \(\sigma_{DM}\) leads to greater offsets between the centers of the galaxy cluster and the dark matter distribution, supporting related theoretical predictions.

\begin{figure}[h]
    \centering
    \epsscale{0.65}
    \plotone{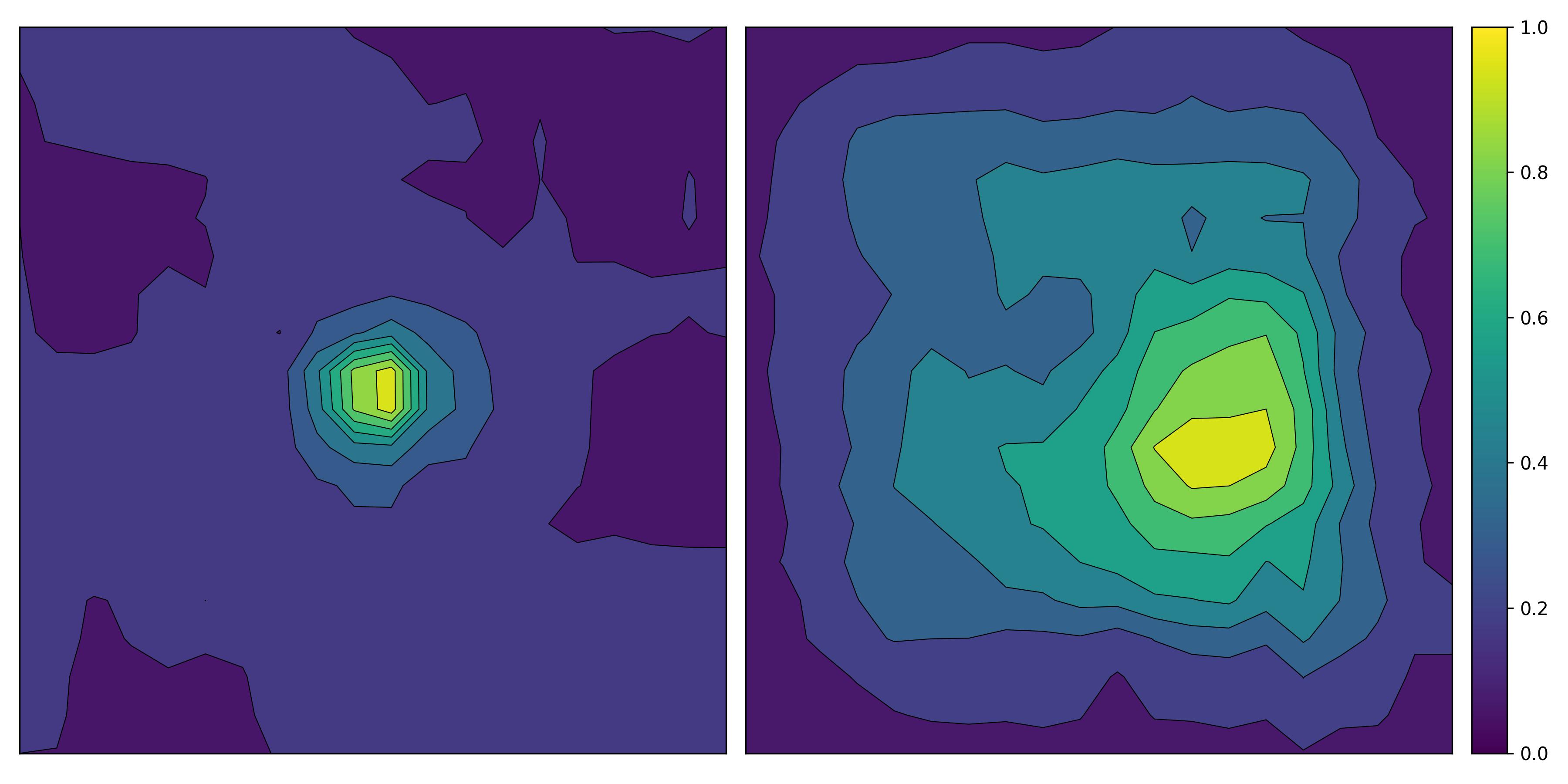}
    \caption{The left and right panels show the difference feature contribution maps for the total and X-ray channels, respectively. These maps are obtained by subtracting the feature contribution of SIDM0.1 from that of SIDM1.0. The color scale bar on the right indicates the normalized difference $h_i$ values. Overlaid contours on each panel highlight the spatial distribution of these differences. In the total channel, we can observe that regions which were previously sparse in contours have become denser, and a compact core has emerged. In the X-ray channel, it can be seen that the offset of the high-contribution core relative to the center is larger in the SIDM1.0 model compared to the SIDM0.1 model.}
    \label{figures:combined_diff}
\end{figure}

In \autoref{figures:combined_diff}, we compare the total maps of SIDM1 and SIDM0.1 by computing the difference between the two maps.
It can be clearly seen that the density increases in areas where the original contour lines have a gentler slope, showing that these regions are becoming more noticeable. 
A high-contribution area also appears here, suggesting that a more defined "core" is forming in what were previously less dense areas. 
Moreover, the central offset between the galaxy cluster and the dark matter halo is much larger in SIDM1 than in SIDM0.1. 
This finding matches the patterns seen in \autoref{figurescombined_total}, supporting the predictions about how the self-interaction cross-section affects dark matter distribution.

The X-ray contribution map shown in \autoref{figures:combined_xray} not only highlights the halo miscentering and mass bias phenomena (as seen in \autoref{figures:combined_diff}), but also reveals that, as the self-interaction cross-section increases, the high X-ray contribution area near the cluster center gradually expands. 
In the early stages of galaxy cluster evolution, SIDM operates under the Long Mean Free Path regime, where thermal conductivity is proportional to the interaction cross-section \citep{10.1093/mnras/stad1705}. 
As the self-interaction cross-section grows, the enhanced thermal conduction works to equalize the temperature within the dark matter halo. 
This allows heat to flow from the hotter outer regions to the cooler core, resulting in a flat central density core during early evolution, which contrasts sharply with the steep density peaks typically seen in CDM models \citep{1996ApJ...462..563N}.

The root-mean-square velocity of material near the core is proportional to $\sigma_{\text{DM}}$ \citep{10.1093/mnras/sts514}, and \cite{Sofue2013} had noted that the heating rate of intracluster material can be considered proportional to this velocity. 
From this, we can infer that the core heating area also scales with $\sigma_{\text{DM}}$, explaining the observed increase in the central X-ray high-contribution region as $\sigma_{\text{DM}}$ increases in \autoref{figures:combined_xray}.

At the end of this section, we can attempt to answer the questions raised earlier. 
In \autoref{figurescombined_total} and \autoref{figures:combined_xray}, it can be observed that the differences between CDM and SIDM0.1 are not sufficiently distinct for reliable classification by the network, whereas the network has strong discriminative ability for SIDM1.0, which aligns with the results of dark matter simulations \citep{Brinckmann:2017uve}. 
From the results in \autoref{fig:Confusion_Matrix}, we observe that the classification accuracy for SIDM0.1 is substantially below the network's average accuracy of 80\%, with the majority of misclassified instances being identified as CDM.
This finding is further corroborated by the blind test on unseen SIDM0.1 clusters (the right panel of \autoref{fig:CKAN_blindtest_result}), where the network underestimates them, interpreting their properties as more akin to CDM.
These results imply that, for CKAN, $\sigma_{\mathrm{DM}}/m = 0.1\,\mathrm{cm}^2/\mathrm{g}$ represents the lower bound for efficient identification of SIDM clusters; in other words, only SIDM clusters with cross-sections exceeding $0.1\,\mathrm{cm}^2/\mathrm{g}$ can be more accurately identified by the network.
In contrast, when testing on unseen SIDM0.3 clusters (the left panel of \autoref{fig:CKAN_blindtest_result}), CKAN tends to overestimate their cross-sections. 
This suggests that CKAN identifies SIDM0.3–dominated cluster distributions as more characteristic of SIDM1.0 than of CDM.
Therefore, we infer that CKAN is capable of discerning the collisional nature of SIDM with $\sigma_{\mathrm{DM}}/m = 0.3\,\mathrm{cm}^2/\mathrm{g}$, successfully disentangling them from collisionless CDM.

% In summary, for CKAN, we estimate the minimum SIDM cross-section $(\sigma/m)_{th}$ needed to disentangle SIDM from CDM on galaxy cluster scales to fall within the range $0.1\,\mathrm{cm}^2/\mathrm{g}$ to $0.3\,\mathrm{cm}^2/\mathrm{g}$. 

In summary, by CKAN, we estimate that for SIDM in galaxy clusters, the minimum cross-section $(\sigma/m)_{\mathrm{th}}$ required to reliably identify its collisional nature falls between $0.1\,\mathrm{cm}^2/\mathrm{g}$ and $0.3\,\mathrm{cm}^2/\mathrm{g}$. 
When the self-interaction cross-section of SIDM is smaller than $(\sigma/m)_{th}$, CKAN may fail to recognize their collisional nature and thus misclassify some of them as CDM, possibly requiring more prior information for the network to learn (such as baryonic effects, dynamical friction, etc. \citep{10.1093/mnras/stx1831,2018MNRAS.476L..20R}).
This, in turn, explains why CKAN is exceptionally effective at identifying SIDM1.0 clusters---whose cross-sections are well above the $(\sigma/m)_{th}$---achieving a near-perfect accuracy of 98\%, which stands in sharp contrast to the network's 80\% average (see \autoref{fig:Confusion_Matrix}).
Through the interpretability and test results of CKAN, we provide a parameter range for SIDM cross-sections on galaxy cluster scales where disentangling SIDM from CDM becomes difficult, consistent with the constraints provided by \cite{2019MNRAS.488.3646R} and \cite{2017PhDT........36E}. 
We expect that with the accumulation of future simulations and observations, CKAN could potentially provide stronger constraints on the SIDM interaction cross-section and uncover additional dark matter features.

\section{Making the inputs observationally realistic}\label{sec:4}

In this section, we assess CKAN’s robustness to realistic observational conditions by introducing noise modeled on two telescope specifications into the input data.
The parameters for calculating noise include: a weak lensing map based on realistic noise estimates for either 50 (Euclid) or 100 (JWST) galaxies per square arcminute; a constant Chandra-like telescope with an X-ray exposure time of 10 ks; and additional default parameters along with their sources can be found in \citep{harvey2024}.
The detailed methodology for noise simulation is provided in \autoref{noise}.

\begin{figure}[H]
    \centering
    \epsscale{0.9}
    \plotone{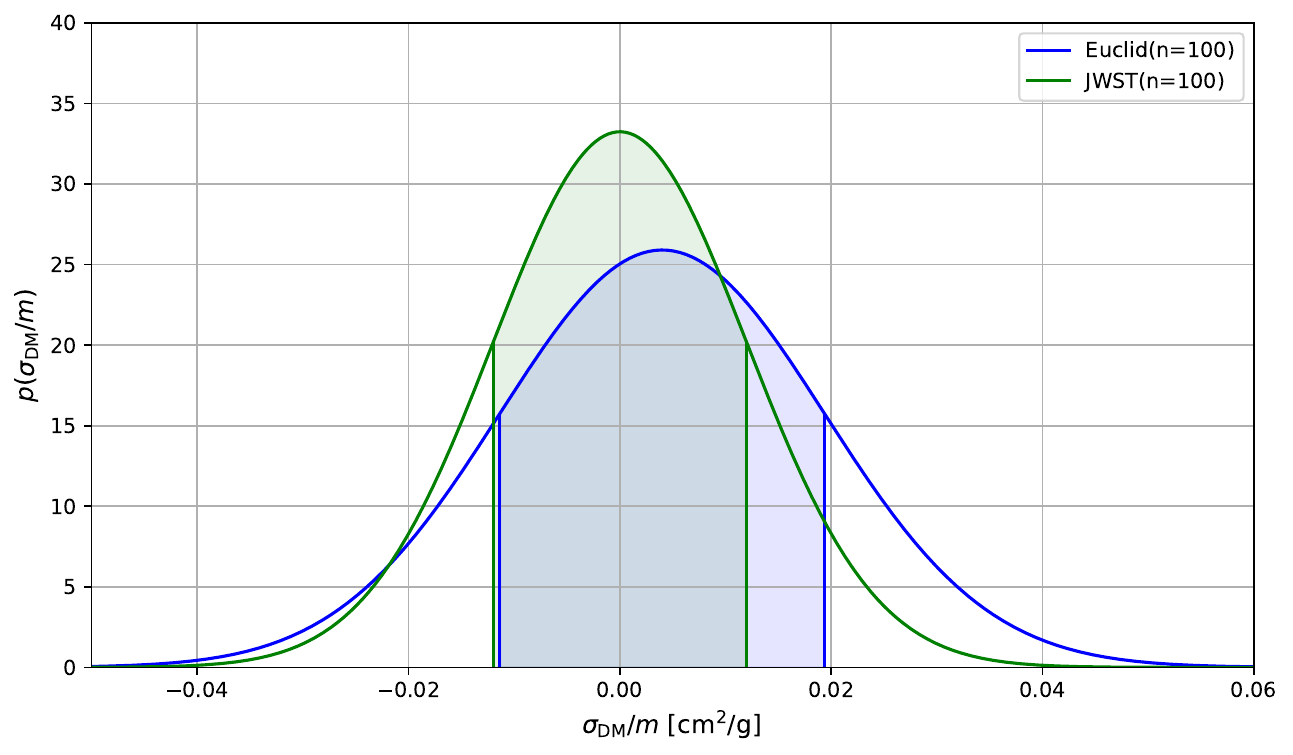}
    \caption{
    Robustness of CKAN under JWST and Euclid noise.
    The network is trained on CDM-low AGN, CDM fiducial AGN and two SIDM models ($\sigma_{\mathrm{DM}}/m = 0.1,1\,\mathrm{cm}^2/\mathrm{g}$), and then tested on the unseen CDM-hi AGN, all these with JWST and Euclid noise. A consistent sample size of n = 100 is used for all tests.}   
    \label{fig:combined_cdm_plot}
\end{figure}

Similar to the approach in Section \ref{Generalization} used for testing generalization ability, we evaluate the trained CKAN on noisy data not included in the training set, thereby assessing its capability to perform on realistic future data.
A comparison with the CNNs \citep{harvey2024} reveals that CKAN, trained on theoretical data, is less effective at identifying noisy data than the larger parameter CNN model (InceptionNet). 
Looking ahead to future telescope technology, we considered a realistic scenario with richer prior information: real noisy data will be included during the training of CKAN as telescopes operate.  
In this scenario, CKAN demonstrates its ability to recognize and generalize noisy data effectively. 
\autoref{fig:combined_cdm_plot} shows the CKAN’s ability to generalize under these conditions, with an expected bias of $\sigma_{\mathrm{DM}}/m < 0.01 \,\mathrm{cm}^2/\mathrm{g}$ on the unseen CDM-hi AGN samples, demonstrating strong robustness.

\section{Conclusion}\label{sec:conclusion}

% In this study, we introduce an efficient CKAN that offers greater interpretability and uses fewer parameters than CNNs for disentangling SIDM and CDM in galaxy clusters.
% By analyzing CKAN's symbolic properties and test results, we find that our key findings align qualitatively or semi-quantitatively with existing theoretical predictions \citep{10.1093/mnras/stab1198,2019MNRAS.488.3646R}.
% These findings include the spatial distribution of the identified features and our estimate that for SIDM in galaxy clusters, the minimum cross-section $(\sigma/m)_{\mathrm{th}}$ required to reliably identify its collisional nature falls between $0.1\,\mathrm{cm}^2/\mathrm{g}$ and $0.3\,\mathrm{cm}^2/\mathrm{g}$.

% Our results show that by using learnable B-spline functions instead of fixed activation functions, CKAN can be built with fewer parameters, enabling simpler explanations of how these networks arrive at their outputs and highlighting important physical patterns from the BAHAMAS-SIDM simulations. 
% We further propose a symbolic approach to interpret these patterns, which not only helps researchers better understand the physical mechanisms behind the data but also builds trust in neural networks. 
% Ultimately, our findings demonstrate CKAN’s promise for next generation telescope surveys and its potential to advance AI‐driven astrophysical analysis, opening new avenues for discovery.

In this study, we present an efficient, interpretable CKAN for disentangling SIDM from CDM in galaxy clusters. 
Using CKAN’s feature attributions and symbolic analysis, we reveal two physically meaningful signatures: (i) the offset between the dark matter distribution center and the galaxy cluster center increases with cross-section, qualitatively consistent with theoretical expectations \citep{10.1093/mnras/stab1198}; and (ii) the growth and flattening of an X-ray–bright, thermally heated central core as cross-section rises, qualitatively matching the trend predicted by \cite{10.1093/mnras/stad1705}.
Combining interpretability analyses with unseen test results, we estimate that for SIDM in galaxy clusters, the minimum cross-section $(\sigma/m)_{\mathrm{th}}$ required to reliably identify its collisional nature falls between $0.1\,\mathrm{cm}^2/\mathrm{g}$ and $0.3\,\mathrm{cm}^2/\mathrm{g}$, broadly consistent with the constraints provided by \cite{2019MNRAS.488.3646R} and \cite{2017PhDT........36E}.

Methodologically, replacing fixed activations with learnable B-spline bases yields a compact network with fewer parameters, enabling simpler explanations of how these networks arrive at their outputs and clarifying the mechanisms behind the classifier’s choices.
We further propose a symbolic approach to interpret these mechanisms, thereby highlighting important physical patterns in the BAHAMAS–SIDM simulations and building trust in neural networks while helping researchers better understand the physics underlying the data.
Together, these results indicate that CKAN is well suited for next-generation survey data and can help advance trustworthy, AI-driven astrophysical inference in cluster dark matter studies.

Although CKAN performs well in parameter efficiency, generalization, and interpretability, its performance on data with observational noise is weaker than that of a CNN when both are trained on noiseless simulated data.
Fortunately, this issue is expected to improve as forthcoming telescope surveys.
Moving forward, we will (i) enrich CKAN’s training sets with broader cross-section variations and relevant physical priors; (ii) integrate higher-resolution simulated or observational data into the learning process; (iii) undertake a thorough analysis of CKAN’s symbolic interpretability; and (iv) extend the framework to other astrophysical scenarios.

\begin{acknowledgments}

This work was supported by National Key R\&D Program of China No.2021YFC2203501. The research was also partly supported by the Operation, Maintenance and Upgrading Fund for Astronomical Telescopes and Facility Instruments, budgeted from the Ministry of Finance of China (MOF) and administrated by the Chinese Academy of Sciences (CAS), and the Scientific Instrument Developing Project of the Chinese Academy of Sciences, Grant No. PTYQ2022YZZD01.

We thank David Harvey for providing the dataset for this study. We would also like to thank Xuwei Zhang, Ziming Liu, Junda Zhou and Yining Song for helpful discussions on topics related to this work.

\end{acknowledgments}

% %% To help institutions obtain information on the effectiveness of their 
% %% telescopes the AAS Journals has created a group of keywords for telescope 
% %% facilities.
% %
% %% Following the acknowledgments section, use the following syntax and the
% %% \facility{} or \facilities{} macros to list the keywords of facilities used 
% %% in the research for the paper.  Each keyword is check against the master 
% %% list during copy editing.  Individual instruments can be provided in 
% %% parentheses, after the keyword, but they are not verified.

% \vspace{5mm}
% \facilities{HST(STIS), Swift(XRT and UVOT), AAVSO, CTIO:1.3m,

% %% Similar to \facility{}, there is the optional \software command to allow 
% %% authors a place to specify which programs were used during the creation of 
% %% the manuscript. Authors should list each code and include either a
% %% citation or url to the code inside ()s when available.

\software{NumPy \citep{harris2020array},
          Astropy \citep{astropy},
          SciPy \citep{2020SciPy-NMeth},
          SymPy \citep{10.7717/peerj-cs.103},
          Matplotlib \citep{Hunter:2007},
          PyTorch \citep{paszke2019pytorchimperativestylehighperformance},  
          KANs \citep{liu2024kankolmogorovarnoldnetworks}.
          }

% %% Appendix material should be preceded with a single \appendix command.
% %% There should be a \section command for each appendix. Mark appendix
% %% subsections with the same markup you use in the main body of the paper.

% %% Each Appendix (indicated with \section) will be lettered A, B, C, etc.
% %% The equation counter will reset when it encounters the \appendix
% %% command and will number appendix equations (A1), (A2), etc. The
% %% Figure and Table counter will not reset.

\appendix

\section{Probability calculation}
\label{calculate:pred}

For the $i$-th galaxy cluster, the probability of the self-interacting dark matter cross-section $\sigma_{\text{DM}}/m$ given the model $X$ can be expressed as:
\begin{equation}
p_i(\sigma_{\text{DM}}/m | X) = e^{X_i},
\end{equation}
where $e^{X_i}$ represents the exponential value of the $i$-th galaxy cluster output by model $X$, interpreted as the probability that the cluster belongs to a specific dark matter cross-section value.

For the combined estimation of multiple clusters, the cross-section estimate for each cluster can be viewed as its probability density function (PDF). Therefore, the combined PDF for multiple clusters can be represented by the product of these individual PDFs:
\begin{equation}
p(\sigma_{\text{DM}}/m) = \prod_{i=1}^{n} p_i(\sigma),
\end{equation}
where $n$ represents the number of galaxy clusters.

The final estimate of the dark matter cross-section can be obtained by calculating the expected value from the resulting probability distribution:

\begin{equation}
\hat{\sigma}_{\text{DM}}/m = \sum p(\sigma_{\text{DM}}/m) \cdot \sigma_{\text{DM}}/m.
\end{equation}

\section{Feature Contribution Map Calculation}
\label{calculate:feature:map}

Starting from the two-dimensional distribution $n \times n$ matrix $M$, the CKAN maps $M$ into a probability vector $v$ within a three-dimensional probability distribution space. The projections of $v$ onto the three axes correspond to the probability amplitudes for each of the three dark matter models, given by:
\begin{equation}
F(M) = v=
\begin{pmatrix}
p_0 \\
p_{0.1} \\
p_{1.0}
\end{pmatrix}.
\end{equation}

Here, $p_0$, $p_{0.1}$, and $p_{1.0}$ denote the probabilities assigned by the CKANs network that the input distribution matrix $M$ corresponds to one of the three dark matter models ($\sigma_\mathrm{DM}/m = 0$, $0.1$, $1.0 \, \mathrm{cm}^2/\mathrm{g}$), respectively.

% \begin{equation}
% \left\{
% \begin{aligned}
% p_0 = P\{\sigma_{\mathrm{DM}}/m = 0.0\mathrm{cm}^2/\mathrm{g}\} &=  f_0(x_1, x_2, \ldots, x_{n^2}) \\
% p_{0.1} = P\{\sigma_{\mathrm{DM}}/m = 0.1\mathrm{cm}^2/\mathrm{g}\} &=  f_{0.1}(x_1, x_2, \ldots, x_{n^2}) \\
% p_{1.0} = P\{\sigma_{\mathrm{DM}}/m = 1.0\mathrm{cm}^2/\mathrm{g}\} &=  f_{1.0}(x_1, x_2, \ldots, x_{n^2}) 
% \end{aligned}
% \right.
% \end{equation}

Since the final fully connected KANs layer corresponds to $p_0$, $p_{0.1}$, and $p_{1.0}$, we can obtain the mapping from the matrix elements to the classification probabilities of the three dark matter models:
\begin{equation}
\left\{
\begin{aligned}
&p_0  =  f_0(x_1, x_2, \ldots, x_{n^2}) \\
&p_{0.1} =  f_{1}(x_1, x_2, \ldots, x_{n^2}) \\
&p_{1.0} =  f_{2}(x_1, x_2, \ldots, x_{n^2}) 
\end{aligned}
\right. .
\end{equation}

\textcolor{black}{In this context, \{$f_0$, $f_{1}$,$f_{2}$\} represents the mapping constructed by the CKANs, which maps each point in the input distribution matrix \( M \) to the output probability \{$p_0$, $p_{0.1}$,$p_{1.0}$\}.}

Finally, we compute the absolute value of the partial derivatives as the contribution measure $h_{i}$, to quantify the contribution of $x_i$ to the final classification result:
\begin{equation}
h_{i} = \left|{\frac{\partial f_k}{\partial x_i}}\right|, \quad k = 0, 1, 2 \ .
\end{equation}

\textcolor{black}{\section{Noise Simulation}\label{noise}}

\subsection{Noise Simulation for Stellar Mass Maps}

Stellar mass maps are often used to describe the distribution of stars within celestial bodies. However, in real observations, stellar mass maps are not obtained directly but are inferred through other methods such as luminosity models or stellar population synthesis models. Importantly, in real-world observations, stellar mass maps inherently contain noise due to limitations in observational instruments and the complex distribution of astronomical objects.

To simulate realistic noise, we incorporated metadata as the stellar channel instead of using the original maps (e.g., X-ray concentration of galaxy clusters, redshift, and properties of the Brightest Cluster Galaxy (BCG)).
These metadata were incorporated into the input of CKAN to make the input data more representative of actual observational conditions. Specifically, noise is introduced through the integration of these external datasets and physical models, a method that has been demonstrated as effective in Harvey’s work \citep{harvey2024}.

\subsection{Noise Simulation for Total Mass Maps}

Total mass maps are derived through gravitational lensing to estimate the mass distribution within celestial systems. For these maps, noise arises from multiple sources, including astronomical background noise, measurement errors, and instrument-related uncertainties.

In our simulation of total Mass Maps, we utilized weak lensing data---a method that does not require parametric modeling and is particularly suited for studying the effects of dark matter, including self-interacting dark matter. 
To simulate the weak lensing process, we first convert mass maps into convergence (\(\kappa\)) maps:

\textcolor{black}{\begin{equation}
\kappa = \frac{\Sigma}{\Sigma_{\rm crit}}.
\end{equation}}

We assume a single lens plane at $z_{\ell}=0.3$ and a single source plane at $z_{s}=2.0$, such that all background galaxies effectively lie at $z_{s}$. 
Here, $\Sigma$ denotes the projected surface mass density, and $D_{\rm ol}$, $D_{\rm ls}$, and $D_{\rm os}$ are the angular diameter distances from the observer to the lens, from the lens to the source, and from the observer to the source, respectively.
The corresponding critical surface density is then given by:

\textcolor{black}{\begin{equation}
    \Sigma_{\rm crit} = \frac{c^2}{4 \pi G} \frac{D_{\rm os}}{D_{\rm ol} D_{\rm ls}}.
\end{equation}}

Next, we apply the Kaiser-Squires method to invert the map and determine the true gravitational shear. This process introduces uncertainties through simulated noise, particularly by adding intrinsic shape noise and other measurement biases when measuring the shear. Gravitational shear, caused by the weak lensing effect, is used to infer the total mass distribution by analyzing distortions in the shapes of celestial objects. 
The complex shear components \(e_{1}\) and \(e_{2}\) are obtained directly from the convergence field \(\kappa\) by applying the Kaiser–Squires inversion \citep{KS93}:

\begin{equation}
(e_{1},e_{2}) \;=\; \mathrm{ks93inv}\!\bigl[\kappa,\,0\bigr] .
\end{equation}

The resulting shear data were then perturbed with Gaussian-distributed shape noise, where the standard deviation of the noise was determined by parameters related to the density of source galaxies. The noise was added using the following equations:

\textcolor{black}{\begin{equation}
e_1' = e_1 + \mathcal{N}(0, \sigma_{e}),
\end{equation}}

\textcolor{black}{\begin{equation}
e_2' = e_2 + \mathcal{N}(0, \sigma_{e}).
\end{equation}}

Here $\mathcal{N}(0,\sigma_e)$ denotes a zero-mean Gaussian random noise with standard deviation $\sigma_e$, where the per-component shape-noise dispersion is given by:
\begin{equation}
\sigma_{e}
= \frac{\ell_{\rm disp}}{\sqrt{2N_{\rm gal}}}.
\end{equation}

The intrinsic ellipticity dispersion $\ell_{\rm disp}=0.3$. \(N_{\text{gal}}\) is the number of galaxies per unit area, given by:
\begin{equation}
N_{\rm gal}
= n_{\rm gal}
  \bigl(\theta_{\rm pix}\bigr)^{2}
\quad\text{with}\quad
\theta_{\rm pix} = \frac{k_{\rm pc/pix}}{D_A(z_\ell)\times(\pi/10800)}.
\end{equation}

Here we assume a continuous, uniform background density $n_{\rm gal}=100\,\mathrm{arcmin^{-2}}$ for JWST and $50\,\mathrm{arcmin^{-2}}$ for Euclid, $k_{\rm pc/pix}=20\,\mathrm{kpc/pix}$ and $D_A(z_\ell)$ denotes the angular diameter distance from the observer to the lens plane at redshift $z_\ell =0.3$  \citep{harvey2024}.

\textcolor{black}{To simulate measurement bias, we introduced shear biases, represented as additive and multiplicative biases for shear components \(e_1'\) and \(e_2'\). The modifications were applied as follows:}
\begin{equation}
e_{i}'' = c_{i} + \bigl(1 + m_{i}\bigr)\,e_{i}' \,\,\, ,\quad(i=1,2),
\end{equation}
with default bias parameters $c_{1}=c_{2}=0$ and $m_{1}=m_{2}=0$.

Finally, we normalized the computed \(\kappa_{E}\) values after noise simulation to compress the intensity range of the images to [0, 1], using the following operation:

\begin{equation}
\kappa_{E} = \mathrm{ks93}\!\bigl[e_{1}'',\,e_{2}''\bigr],
\end{equation}
\begin{equation}
\kappa_{E}^{\rm norm}
= \frac{\kappa_{E} - \min(\kappa_{E}) + \langle\kappa\rangle}
       {\max\bigl[\kappa_{E} - \min(\kappa_{E}) + \langle\kappa\rangle\bigr]}.
\end{equation}

\subsection{Noise Simulation for X-ray Emission Maps}

\textcolor{black}{X-ray emission maps are typically observed through X-ray telescopes such as the Chandra X-ray Observatory and assume a specific spectral energy distribution. In real observations, these maps are affected by noise, primarily arising from sampling noise of the instrument and the detection precision of the detectors. We assume an average X-ray photon energy of 2 keV, with the photons having the same redshift as the lens (\(z_l = 0.5\)).}

\textcolor{black}{Noise is introduced based on the following assumptions:}

\textcolor{black}{\begin{enumerate}
    \item Poisson sampling: the X-ray emission intensity at each pixel follows a Poisson distribution, representing the number of detected X-ray events per unit time. Due to the statistical nature of the detector, noise results from random fluctuations in the independent observation counts within each pixel.
    \item Exposure time: the noise level depends not only on the intrinsic intensity of the source but also on the simulated exposure time. A longer exposure time increases \(\lambda\) (expected photon counts), smoothing out Poisson noise and reducing the impact of errors on the model.
\end{enumerate}}

\textcolor{black}{The specific calculation steps are as follows:}

First, based on the input data's dimensionless X-ray emission and X-ray norm, we compute the X-ray emission intensity at each pixel (units: \(erg/s/cm^2/arcmin^2\)):

\begin{equation}
E_{ij} = F_{0}\,\epsilon_{ij}\,T_{\rm exp}\,\Delta\Omega_{\rm pix}\,A_{\rm tel}.
\end{equation}

Where \(F_{0}\) is the X‐ray normalization factor (units: \(\mathrm{erg\,s^{-1}\,cm^{-2}\,arcmin^{-2}}\)), \(\epsilon_{ij}\) is the dimensionless emission coefficient at pixel \((i,j)\) , \(T_{\rm exp}\) is the exposure time in seconds, \(\Delta\Omega_{\rm pix}\) is the solid angle per pixel in arcmin\(^2\), and \(A_{\rm tel}\) is the telescope collecting area in cm\(^2\). 
Assuming a photon energy of 2 keV, we can calculate the number of photons per pixel using:

\textcolor{black}{\begin{equation}
N_{\text{p}} = \frac{E_{ij}}{2 \cdot \text{keV}}.
\end{equation}}

\textcolor{black}{Next, we simulate the photon count for each pixel using a Poisson distribution, generating random observed counts for the expected count \(N_{\text{p}}\) as follows:}

\textcolor{black}{\begin{equation}
P(N) = \frac{\lambda^N e^{-N_{\text{p}}}}{N!}.
\end{equation}}

\textcolor{black}{In addition to signal noise, we also introduce background noise (e.g., from astronomical background radiation). Background noise also follows a Poisson distribution, consistent with the noise characteristics of the signal, as background radiation occurs randomly and independently of other pixels or observations. The background noise for each pixel is described by:}

\textcolor{black}{\begin{equation}
P(N_{\text{bkgd}}) = \frac{\lambda_{\text{bkgd}}^{N_{\text{bkgd}}} e^{-\lambda_{\text{bkgd}}}}{N_{\text{bkgd}}!}.
\end{equation}}

Where \(N_{\text{bkgd}}\) is the pixel's background noise count, and \(\lambda_{\text{bkgd}}\) is the expected count for background noise, typically depending on the intensity of the astronomical background radiation and the sensitivity of the instrument (with different parameters used depending on the telescope selected) \citep{harvey2024}. By combining the simulated photon counts and background noise, we obtain the final simulated X-ray observation map. Similar to the total mass maps, we normalize this map.

\bibliography{sample631}{}
\bibliographystyle{aasjournal}



\end{document}